\documentclass[a4paper,11pt]{article}
\usepackage{pos}

\title{Anomalies, $\eta$ , $\eta$' as keys to glueballs.}
\ShortTitle{Eta(prime)s reveal glueballs }

\author*[a]{Jean-Marie Fr\`{e}re  }

\affiliation[a]{Theoretical Physics CP225, Universit\'{e} Libre de Bruxelles,\\
  blvd du Triomphe , 1050 Brussels, Belgium, \\ and Brout-Englert-Lema\^itre Center, Brussels}

\emailAdd{frere@ulb.be}

\abstract{Glueballs are the most straightforward prediction of QCD, yet while they have likely been produced, none has been unequivocally identified. We pursue a backdoor approach through anomalies, and singularly the $\eta$ and $\eta$' which  brings light to this irritating situation.
In particular, we advocate to consider the full decay chain  $J/\psi \rightarrow X \gamma , X \rightarrow \eta \eta'$ (into glue-rich states followed by glue-rich decays). We also suggest new BES III searches, namely for the $\pi_1$ into $\eta(') \pi^0$, (this would be the partner of their recently observed $\eta_1(1855)$). Another useful investigation would be for other channels (or semi-inclusive) $f_0 (1500)$ decays (see last section)}

\FullConference{%
  Corfu Summer Institute 2022 "School and Workshops on Elementary Particle Physics and Gravity",\\
  28 August - 1 October, 2022\\
  Corfu, Greece}

\begin{document}
\maketitle
\section{Introduction}

While glueballs (colour neutral bound states of 2 or more gluons) are an obvious and compelling prediction of Quantum Chromodynamics (QCD), none have this far been identified with certainty.
In a way, most physicists have ceased searching for them, satisfying themselves that they have certainly been produced, but are lost in a complex forest of quark states, with  which they mix, pointing in particular at the total number of $0^{++}$ or $0^{+-}$ states between 1 and 2 Gev.
More conspicuous would however be "exotics", namely mesons with the "wrong" charge-parity assignment with respect to the na\"{\i}ve Quark Model as a result of the presence of an extra gluon in their composition; however here also, an ambiguity may exist with 4-quark states, a pair of quarks simulating a gluon.

When glueballs searches have thus become a backwater, it is a bit shocking that such a fundamental test of QCD (and, accessorily , lattice calculations) is neglected.
Fortunately, some long-overdue systematic searches combining gluon-rich production and gluon-rich decays have now been realized, thanks on the one hand to the identification of the $\eta, \eta '$ as possible markers, and on the other hand, to the coming in full operation of BESIII in China, allowing for a never-reached before resolution in the study of the $J/\psi$ radiative decays.

\section{Are glueball decays flavor blind?}
For lack of trustable calculations (we are deep in the domain of really strong forces, far away from asymptotic freedom) has led to a number of qualitative assumptions about the decays of glueballs.
\begin{itemize}
  \item the straightforward assumption is that decays are "flavor-blind", since pure glue ignores this quantum number, with the mitigation of phase space.
  \item very quickly, qualifications and possible deviations were suggested: wave function overlap (a quantum mechanical approach) would suggest that heavier particles, closer to the 1-2 GeV mass of the glueballs would be preferred , say $K$ pairs over $\pi's$.
  \item this tendency is also advocated by the supposition that  2 gluons cannot couple to light quarks, since the decay in a pseudoscalar meson would necessitate a mass insertion
  \item it should however be noted that the above argument does not really apply in general cases, since it is well-known that quantum anomalies precisely relate the divergence of the axial current or light quarks to pure gluons $\widetilde{G^{\mu \nu}} G_{\mu \nu}$ : this is precisely the term which appears in the evaluation of $\eta , \eta'$ states.  A more detailed discussion can be found in ref. \cite{Frere:2015xxa}.
  \item the above arguments are mostly limited to 2-body decays into pseudoscalars, but it is not absurd to assume that flavor-neutral states would decay predominantly by 2 light $\sigma$ (the very broad partner of the $\pi$, only seen by partial wave interference), those particles decaying in turn in pion pairs (thus at least 4 $\pi$)
\end{itemize}

\section{Enter the $\eta$ and $\eta'$}

Some early hints of modern glueballs searches came early (early 1980's) with the GAMS experiment \cite{Serpukhov-Brussels-AnnecyLAPP:1983xdr},  \cite{Serpukhov-Brussels-AnnecyLAPP:1983jxn}, which investigated a "gluon-rich" (central production) environment using a lead glass wall, mostly sensitive to photons. This had the effect of accessing the $\eta$ and $\eta'$ decay modes, and resulted in at least one glueball candidate (which they called G(1590) , now the $f_0$ (1500);the discrepancy in mass is most probably due to the limited phase space in the $\eta - \eta'$ channel, which leads to severe distortion).

Interestingly, the same experiment (which was conducted in parallel at CERN and Serpukhov) led to the discovery of an exotic candidate, then called M(1406), now probably $\pi_1$(1400) (also sometimes called $\widetilde{\rho }$ , with exotic parity $1^{-+}$ impossible for a quark-antiquark meson), seen decaying into $\eta(') \pi$.
In both cases, the role of the $\eta'$ meson was key, as pointed out by Gherstein et al.\cite{Gherstein}. A special connection between glue and $\eta'$ was indeed known both from theoretical considerations (QCD anomaly) and the observation of the large $J/\psi \rightarrow \eta' \gamma$ branching ratio.

\begin{figure}[h]
\begin{center}
\includegraphics [width=10cm]{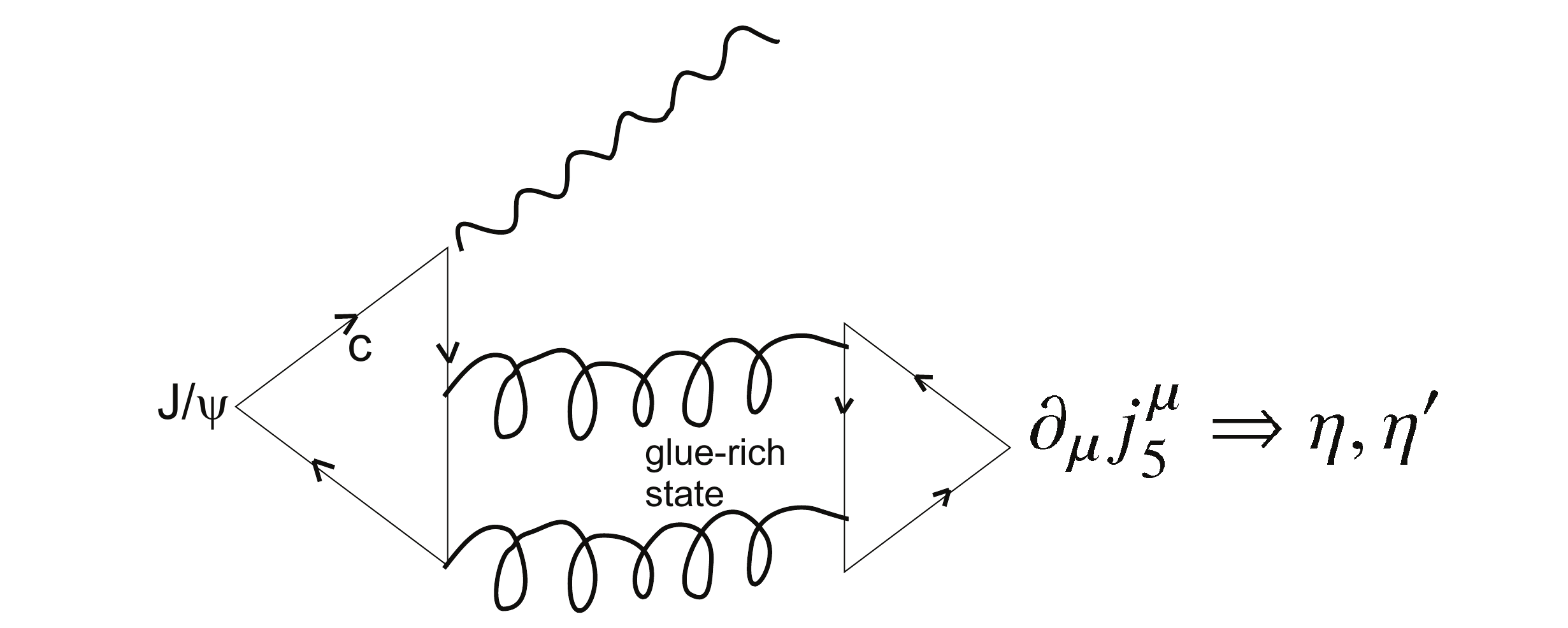}
\caption{J/$\psi$ radiative decay as a glue-rich source coupled to the $\eta$('); notice that no mass insertion is needed }
\label{JradDecay}
\end{center}
\end{figure}

It must be noted that \emph{\textbf{no quark mass insertion is needed in this case}}, due to the anomaly.

\textbf{We have developed a complete formalism to handle the radiative decays of the type V->P $\gamma$ or P->V $\gamma$ on the basis of QCD anomalies,}
\cite{Ball:1995zv}, \cite{Akhoury:1987ed}, further refined in \cite{Escribano:2005qq}, which lead to a successful description of those processes without recourse to additional parameters. The larger decay of J/$\psi$ into $\eta' \gamma$ than $\eta \gamma$ despite an unfavorable phase space is then put in relation to the greater admixture of the anomaly contribution in the physical $\eta '$.

For $J/\psi$ decays, the emission of a photon leaves a glue-rich situation, which eventually couples to the $\eta (')$.

This is thus a different "glue-rich" production source, in addition to the central production. While intensity-limited by the production mechanism, it offers a  much "cleaner" situation to look  for glueball states.

Unfortunately, the J/$\psi$ radiative decays into neutral mesons have proven very difficult experimentally, and the results of a search for glueballs were confusing at first.  While SLAC ceased this activity, more powerful machines preferred to turn to the heavier B mesons. It was not until BES III in Beijing started taking data in recent years that difficult channels could be studied with (impressive) accuracy.

\begin{figure}[h]
\begin{center}
\includegraphics [width=10cm]{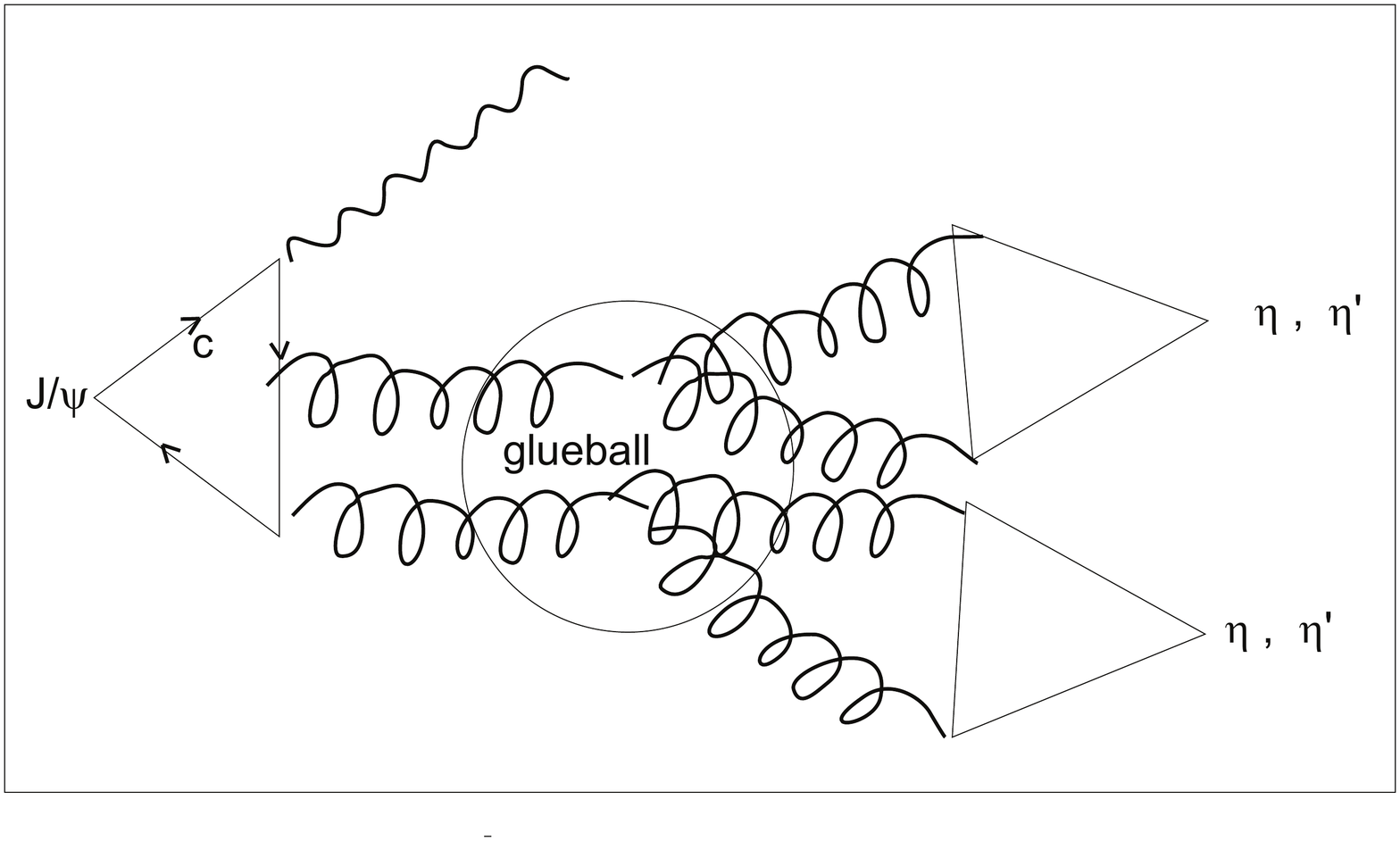}
\caption{J/$\psi$ radiative decay as a glue-rich source leading to Glueball formation and decay, here in the $\eta$ $ \eta'$ channels }
\label{JradDecay}
\end{center}
\end{figure}

\section{Who are the Glueballs? }

\begin{figure}[h]
\begin{center}
\includegraphics [width=12cm]{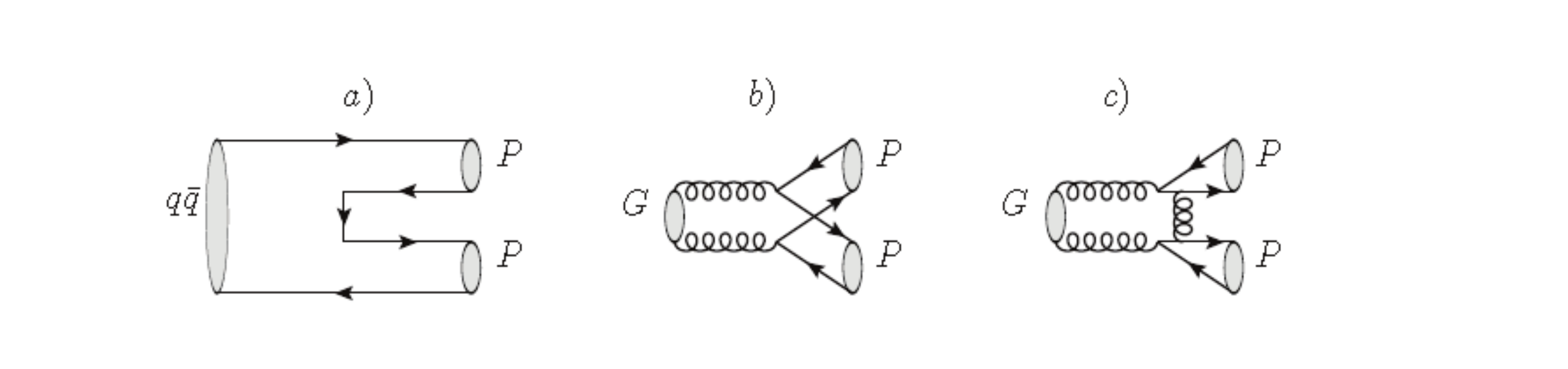}
\caption{3 graphs (in addition to figure \ref{JradDecay} describing possible decay topologies, from ref. \cite{Frere:2015xxa} }
\label{3graphs}
\end{center}
\end{figure}

Many attempts have been made to identify the glueballs in the 1-2 GeV spectroscopy of mesons, with approaches including various elements (in particular, including or excluding the anomaly-mediated channels).
A popular approach (see references in \cite{Frere:2015xxa}) attempts to parametrize the various contributions according to topologies assumed from the schematic mechanisms described, for instance in fig. \ref{3graphs}, or possibly in the left part of fig. \ref{JradDecay}.
While tempting, this approach should be taken with a grain of salt. Due to strong interactions, these diagrams are just oversimplified evocations of processes, and in no way on the level of Feynman graphs!. Each of them should, be enriched with a large number of QCD extra contributions, the phases are unknown, ...

\begin{figure}[h]
\begin{center}
\includegraphics [width=12cm]{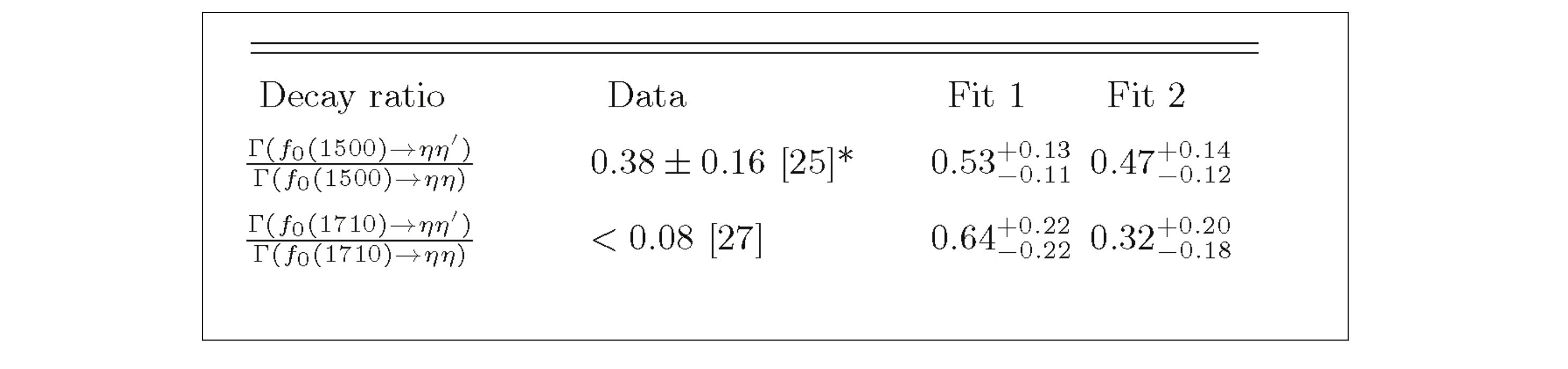}
\caption{Our previous fit to the  $f_0 (1710)$ and  $f_0 (1500)$
 decays into $\eta \eta'$, allowing for the severe phase space suppression of the latter; the brackets refer to the data available at the time  \cite{Frere:2015xxa}  and the whole fit is now superseded bu BES III data}
\label{oldfit}
\end{center}
\end{figure}

As others, we have tried this approach a few years ago, and tried to compensate the insufficient theoretical basis with a long list of processes to be fitted. On the basis of this analysis, we tried 2 fits, one without assuming the often advocated chiral suppression (against which we already listed arguments)  and the more traditional one for comparison.

Both fits (in line with the then tendency) tended to present the $f_0 (1710)$ as a mostly-glueball state and the $f_0 (1500)$ mostly as a "normal" meson. As part of the fit, we had come with the fit/prediction for the decay of these mesons into $\eta \eta'$ shown in figure \ref{oldfit}  (the reference in brackets refer to the then-available data, see the original paper).

These attempts certainly suggested that the $f_0 (1710)$ would be observed in the $\eta \eta'$ channels when the BES III data would become available!

\section{The new results from BES III change the perspective}

We have finally come close to the holy grail!

 BES III recentlly published a partial wave analysis of the radiative $J/Psi$ decays complete with the $\eta$ modes.  This is precisely the tool which has been missing for tens or years (more from lack of interest than from lack of feasibility for other facilities!).
\cite{BESIII:2022iwi}

And it comes with a surprise (we will deal with another surprise in the "exotics" in a later section).

Namely, the $f_0 (1710)$ is simply not seen in the full decay chain $J/\psi \rightarrow X \gamma , X \rightarrow \eta \eta'$,
where it should dominate the $f_0 (1500)$.
Even accepting a confusion with the $f_0 (1810)$ (remember that the "mass" in the tables is not always the latest
measurement, and that partial wave analysis may be affected by interferences, phase space distortions), we see that it is suppressed in the ration of 0.007/3.05.

\begin{figure}[h]
\begin{center}
\includegraphics [width=14cm]{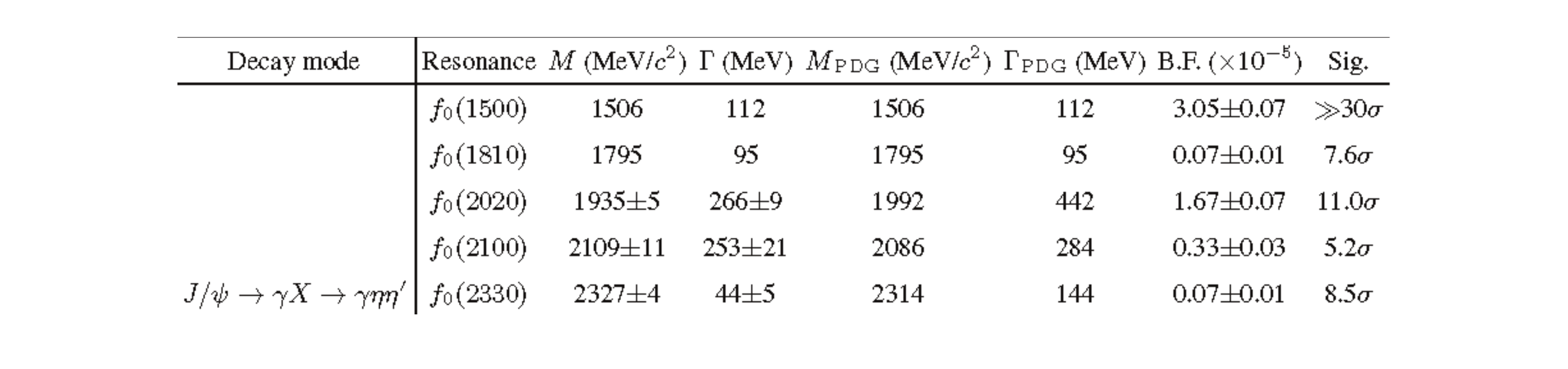}
\caption{The partial wave analysis of the full decay chain $J/\Psi \rightarrow X \gamma , X \rightarrow \eta \eta'$, from table II of \cite{BESIII:2022iwi} }
\label{BESeta}
\end{center}
\end{figure}

See however also a comprehensive analysis by ref \cite{Chen:2022asf}.

\textbf{This suggests another attitude, and strongly points to the combined chain of decay as a prominent test for glueballs.}

Ironically, this most recent results brings us back to the initial observations by GAMS, with their G(1590) , now identified as the $f_0$ (1500) as a leading candidate. Here also, the shift in estimated mass (from the very early and sparse observations) may be understood by the severe distortion of the phase space (the $\eta \eta'$ channel being at the nominal threshold, and only possible due to the state width).

This of course prompts us to look back into the unusual properties of the $f_0 (1500)$, but also to check what the BES III data tell us about the other "weird" observation of GAMS, namely the $1^{-+}$ exotic. This will occupy the next 2 sections.

\section{The very unusual decays of the $f_0 (1500)$ particle}

It is worth  to have a new peek into the PDG tables to learn more about this state, which is back in the spotlight. \begin{figure}[h]
\begin{center}
\includegraphics [width=18cm] {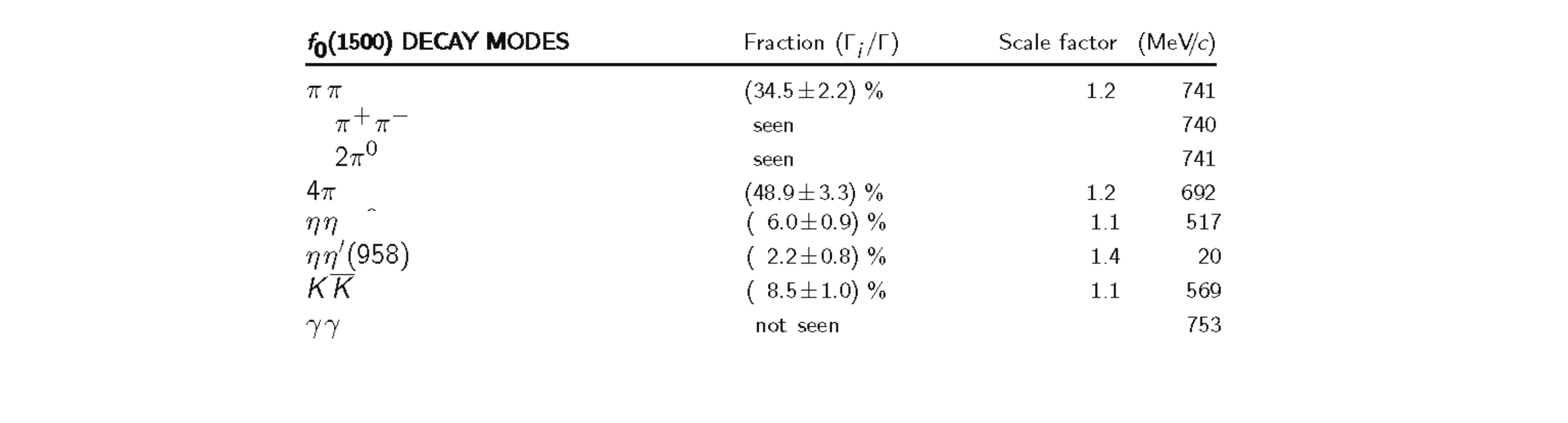}
\caption{The main decay modes of the $f_0 (1500$ according to Particle Data Group \cite{PDG2022}}
\label{PDG}
\end{center}
\end{figure}
In this table, we note that, as already mentioned, the $\eta \eta'$ mode, while at threshold, is very significant (and would dominate the $\eta \eta$ mode after phase space corrections), but also that the main observed decay mode is in $ 4 \pi$, overwhelming the $2 \pi$ channel.
An interesting question remains : there may very well be other, not yet observed modes (multi pions), for which the analysis have not been performed or would prove impractical. In fact, a semi-inclusive  search for $J/\psi \rightarrow \gamma X(1500)$ where $X$ is any recoiling state with invariant mass in the range could be instructive!
Remaining with the $ 4 \pi$ decay mode, this can certainly suggest a decay through a pair of very wide $0^{++} \  \sigma$
states. Given its quantum numbers, the $\sigma$ can certainly be suspected to mix in the "glueball complex".

\section{Return to the Exotics}

Exotics is the name traditionally given to meson states which cannot be reached by the association of just one quark and an antiquark. Such basic combinations indeed have their parity and charge conjugation directly related to orbital momentum and total spin respectively, namely $ P= (-)^{l+1}, C = (-)^{l+s}$. As a result, a $J^{PC} = 1^{+ -}$ state is forbidden.
If a gluon(or a quark-antiquark pair mimicking a gluon) is added, the selection rule fails.

Such a state was advocated by the GAMS collaboration (see above)\cite{IHEP-Brussels-LosAlamos-AnnecyLAPP:1988iqi}, with a decay into $ \eta(') \pi$, and a mass 1406 MeV, (initially nwmaed M(1406) and probably currently known as $\pi_1 (1400))$).
While it was a clear sign of an exotic, the phenomenological analysis conducted then were not enthusiastic, since prejudice disfavored this precise decay mode -- once again, this was an error due to disregarding axial anomalies, and the special role of the $\eta'$!

We advocated already at that time that the inclusion of the anomaly would on the opposite make it one leading decay mode \cite{Frere:1988ac}.

Little work has gone on this search since, but it is interesting that BESIII has now identified a similar exotic, this time in the closely related $ \eta" \eta$ channel, namely a $\eta_1 (1855$ with $I^G J^{PC} = O^+1^{-+}$.\cite{BESIII:2022iwi}
This time, the theoretical interpretation did not miss the role of the anomaly \cite{Chen:2022qpd}, coming to a conlusion similar to ours in the case of the $\pi_1$.

This comforts us in believing that the $eta$ system, through quantum anomalies, is a key to understanding the glueballs and exotics system!

\section{Conclusions and Work to Do}

Obviously our work in understanding the role of gluons in exotic mesons and glueballs is far from complete.
Some comfort may be found in the fact that a large body of experimental evidence is now becoming available in what
was for a long time a neglected field. Many theorists had probably given up in front of the complexity and the lack
of fresh data, but hope is returning and results are lining up nicely.

Many channels still remain worth exploring. Obviously, the search for the exotic $\pi_1$ in the $\eta' \pi^0$ mode
should be conducted at BES III, confirming the old GAMS result and giving extra support for a partner  their $\eta_1$ state.
Further decay modes (ideally a semi-inclusive study) or the $f_0 (1500)$ would also help better understand this state.

The next step would be to use then better established glueballs by probes for heavier states (assuming that a glueball to glueball + X decay  would be favored).

We may see a new sunrise on glueballs!

\section{Acknowledgements}
This work is supported by IISN (Belgium) , the Brout-Englert-Lema\^{\i}tre Center (Brussels) .
I wish to thank the organizers of the Corfu meetings for the occasion to evoke those issues in beautiful surroundings.


\begin{thebibliography}{99}

\bibitem{Frere:2015xxa}
J.~M.~Fr\`ere and J.~Heeck,
``Scalar glueballs: Constraints from the decays into $\eta$ or $\eta'$,''
Phys. Rev. D \textbf{92} (2015) no.11, 114035
doi:10.1103/PhysRevD.92.114035
[arXiv:1506.04766 [hep-ph]].

\bibitem{Serpukhov-Brussels-AnnecyLAPP:1983xdr}
F.~G.~Binon \textit{et al.} [Serpukhov-Brussels-Annecy(LAPP)],
``G (1590): A Scalar Meson Decaying Into Two eta Mesons,''
Nuovo Cim. A \textbf{78} (1983), 313
CERN-EP/83-97.

\bibitem{Serpukhov-Brussels-AnnecyLAPP:1983jxn}
F.~G.~Binon \textit{et al.} [Serpukhov-Brussels-Annecy(LAPP)],
``Observation of Reaction $\pi^- p \to \eta^\prime \eta n$ and a Search for Glueball,''
Sov. J. Nucl. Phys. \textbf{39} (1984), 526
IFVE-83-204.

\bibitem {Gherstein}
Gershtein, S.S., Likhoded, A.A. and  Prokoshkin, Y.D. G(1590)-Meson and possible characteristic features of a glueball,
Z. Phys. C - Particles and Fields 24, 385 (1984).


\bibitem{IHEP-Brussels-LosAlamos-AnnecyLAPP:1988iqi}
D.~Alde \textit{et al.} [IHEP-Brussels-Los Alamos-Annecy(LAPP)],
``Evidence for a 1-+ Exotic Meson,''
Phys. Lett. B \textbf{205} (1988), 397
doi:10.1016/0370-2693(88)91686-3


\bibitem{Ball:1995zv}
P.~Ball, J.~M.~Frere and M.~Tytgat,
``Phenomenological evidence for the gluon content of eta and eta-prime,''
Phys. Lett. B \textbf{365} (1996), 367-376
doi:10.1016/0370-2693(95)01287-7
[arXiv:hep-ph/9508359 [hep-ph]].

\bibitem{Akhoury:1987ed}
R.~Akhoury and J.~M.~Frere,
Phys. Lett. B \textbf{220} (1989), 258-264
doi:10.1016/0370-2693(89)90048-8

\bibitem{Escribano:2005qq}
R.~Escribano and J.~M.~Frere,
JHEP \textbf{06} (2005), 029
doi:10.1088/1126-6708/2005/06/029
[arXiv:hep-ph/0501072 [hep-ph]].

\bibitem{BESIII:2022iwi}
M.~Ablikim \textit{et al.} [BESIII],
``Partial wave analysis of J/\ensuremath{\psi}\textrightarrow{}\ensuremath{\gamma}\ensuremath{\eta}\ensuremath{\eta}',''
Phys. Rev. D \textbf{106} (2022) no.7, 072012
doi:10.1103/PhysRevD.106.072012
[arXiv:2202.00623 [hep-ex]].

bibitem{PDG2022}
R.L. Workman et al. (Particle Data Group), Prog.Theor.Exp.Phys. 2022, 083C01 (2022)

\bibitem{Frere:1988ac}
J.~M.~Frere and S.~Titard,
``A NEW LOOK AT EXOTIC DECAYS rho-tilde (1-+, I = 1) ---\ensuremath{>} eta-prime pi versus rho pi,''
Phys. Lett. B \textbf{214} (1988), 463-466
doi:10.1016/0370-2693(88)91395-0

\bibitem{Chen:2022qpd}
H.~X.~Chen, N.~Su and S.~L.~Zhu,
`QCD Axial Anomaly Enhances the \ensuremath{\eta}\ensuremath{\eta}' Decay of the Hybrid Candidate \ensuremath{\eta} $_{1}$(1855),''
Chin. Phys. Lett. \textbf{39} (2022) no.5, 051201
doi:10.1088/0256-307X/39/5/051201
[arXiv:2202.04918 [hep-ph]].

\bibitem{Chen:2022asf}
H.~X.~Chen, W.~Chen, X.~Liu, Y.~R.~Liu and S.~L.~Zhu,
``An updated review of the new hadron states,''
Rept. Prog. Phys. \textbf{86} (2023) no.2, 026201
doi:10.1088/1361-6633/aca3b6
[arXiv:2204.02649 [hep-ph]].

\end{thebibliography}
\end{document}